\documentclass[12pt]{article}
\usepackage{graphicx}
\usepackage{bm}
\usepackage[utf8]{inputenc}
\usepackage{lmodern}
\begin{document}
\title{On the relevance of quantum models for plasmas} 
\date{}
\maketitle 
\begin{center}
\large{F. Haas 
\vskip.3cm
Departamento de F{\'i}sica, Universidade Federal do Paran\'a, 81531-990, Curitiba, Paran\'a, Brazil}
   \end{center}
   
   \vskip.3cm
\begin{abstract}{\noindent Recently there has been a claim on the complete irrelevance of quantum modeling for plasmas. We address this subject from basic principles. Physical situations where quantum effects play a decisive role are identified.}
\end{abstract}
\vskip.3cm

Pacs numbers: 67.10.Hk, 52.35.Mw, 52.35.Qz

\vskip.5cm
Recently Vranjes, Pandey and Poedts wrote a series of comments \cite{vpd1, vpd2} on the limited applicability of quantum effects in plasmas, except for atomic and/or nuclear aspects like ionization, calculation of scattering cross sections and so on. In spite of the prompt reply by Shukla and Akbari-Moghanjoughi \cite{Shukla}, besides the existing tho\-rough discussions on the basic theoretical models for quantum phenomena in plasmas \cite{Vladimirov, Haas, Manfredi, Bonitz}, 
we fell it would be still of some interest to have more independent views on the topic. Hence we address the question from first principles, apologizing for the superposition with some well-known arguments. % - at least, arguments that {\it should} be well-known. 

For any collection of particles, one may roughly separate quantum effects in three categories: (a) the class of quantum effects due to particle indistinguishability; (b) the family of phenomena originating from wave-particle duality; (c) the class of quantum effects linked to the intrinsic magnetic moment of the charge carrier dynamics, irrespective of quantum statistical issues. Let us start discussing point (a), with an eye to systems of massive fermions since usually electrons are an essential component of plasmas. In this case, there are two main possible descriptions which are in order, based resp. on classical, Maxwell-Boltzmann (MB) or quantum, Fermi-Dirac (FD) statistics. The MB is an approximation to the FD statistics, valid when there is not too much wave packet overlapping. The condition for the applicability of MB is given by $T \gg T_F$, where $T$ is the thermodynamic temperature and 
\begin{equation}
\label{e1}
T_F = \frac{\hbar^2}{2m\kappa_B} (3\pi^2 n_{0})^{2/3} 
\end{equation}
is the Fermi temperature. In Eq. (\ref{e1}), $\hbar$ is Planck's constant divided by $2\pi$, $m$ is the charge carriers mass (in this text we consider electrons), $\kappa_B$ is Boltzmann's constant and $n_0$ is a reference equilibrium number density. Apart from an irrelevant numerical factor of order unity, from Eq. (\ref{e1}) the sufficiently high temperature condition is found to be equivalent to $n_{0}^{-1/3} \gg \lambda_B$, where 
\begin{equation}
\label{e2}
\lambda_B = \frac{h}{\sqrt{2\pi m\kappa_B T}} 
\end{equation}
is the thermal de Broglie length for a Fermi gas. Therefore, if the mean inter-particle distance is much larger than the typical wave packet size one may disregard the indistinguishability aspects. For simplicity, we are talking about a completely ionized gas; for bound particles some appropriate length scale should replace the thermal de Broglie length as a measure of the wave functions size.

In conclusion, a many-body fermion system deserves a quantum statistics treatment 
whenever it is sufficiently cold or dense. Accordingly we refer to degenerate (FD) and non-degenerate (MB) plasmas. We don't intend to present an exhaustive list of real degenerate plasmas, limiting ourselves to few examples: (i) the plasma appearing in intense laser compression schemes \cite{Mourou}, which nowadays can have large densities of the order $n_0 \simeq 10^{32} m^{-3}$. In this case one need the unlikely elevated temperature $T \gg 10^7 K$ to be allowed to apply MB statistic without committing serious error. One should note that with the increasing laser power entering the multi-petawatt range one can 
even envisage to probe strong quantum aspects of nature, including non\-linear modifications of the Maxwell equations due to 
photon-photon scattering mechanisms. Such avenues, however, can be accessed only through electric field strengths of the order of the Schwinger limit $m^2 c^3/(e \hbar) \simeq 10^{18} V/m$, where $c$ is the speed of light and $e$ the elementary charge. Quantum field theoretical aspects are far outside the scope of most quantum plasma models; (ii) the plasma in compact astrophysical objects like white dwarfs \cite{Kepler} and neutron stars \cite{Samuelsson} where $n_0 \simeq 10^{36} m^{-3}$. In this case FD statistics is necessary unless $T \gg 10^9 K$, which is also unlikely. 

At least one can briefly mention some further quantum plasma examples, one of the most notorious being the degenerate electron gas in metals as remembered in \cite{Shukla}. As remarked in \cite{vpd2}, the properties of such  quantum plasmas have been already discussed decades ago. Also one can consider degenerate plasmas in the interior of giant pla\-nets like Jupiter, plasmas confined in traps at sub-kelvin temperatures, and ultra-cold plasmas (with $T \leq 10^{-3} K$) generated
from Rydberg states. We urge the reader to examination of the Figure 1 of Refs. \cite{Manfredi, Bonitz, Brodin, rmp}, showing density versus temperature diagrams where the MB and FD regions and specific physical systems in both classes are clearly identified. Notice that in \cite{vpd1} a similar density versus temperature is shown, but restricted to densities of the order of $10^{17} m^{-3}$, which is obviously not representative as already pointed out in \cite{Shukla}. Following the comments in \cite{Bonitz}, from more complete diagrams it is easy to realize that the family of Coulomb systems (many-body systems where the electric force plays a dominating role) has grown far beyond the conventional laboratory and space plasmas. In the same sense, the very concept of ``plasma" is con\-ti\-nuous\-ly evolving. In this context, the relevance (or irrelevance) of quantum statistical effects needs a case-by-case analysis in each concrete situation. 

Let us now turn our attention to the second class of quantum effects which can impart the behavior of a many-body system, namely class (b) due to wave-particle duality. This includes the whole manifold of quantum diffraction effects, appearing already in the framework of the Schr\"odinger equation (in the non-relativistic case), without one-to-one relation with the Pauli exclusion principle. For instance, wave function spreading and tunneling are among the eminent undulatory processes which are of recognized relevance in the dynamics of ultra-small semiconductor devices. In this res\-pect, consider the Wigner-Poisson model \cite{Markowich} for electrostatic fields in one spatial dimension,
\begin{eqnarray}
\frac{\partial f}{\partial t}\!\!&+&\!\!
\frac{p}{m}\,\frac{\partial\,f}{\partial\,x} + \frac{ie}{2\pi\hbar^2}\int dp'ds\exp\left(\frac{i(p - p')s}{\hbar}\right) \times \nonumber  \\
\label{e3}
\!\!\!\!&\times&\!\!\!\!\left(\phi\left(x\!+\!\frac{s}{2},t\right)\!-\!\phi\left(x\!-\!\frac{s}{2},t\right)\right)f(x,p',t)\!=\!0 , \\
\label{e4}
\frac{\partial^2\phi}{\partial x^2} &=& \frac{e}{\varepsilon}\left(\int dp'\,f(x,p',t) - n_{0}(x)\right) \,.
\end{eqnarray}
In Eqs. (\ref{e3})-(\ref{e4}), $f = f(x,p,t)$ is the Wigner function, $\phi = \phi(x,t)$ is the scalar potential, $\varepsilon$ is the permittivity and $n_{0}(x)$ is a position-density ionic background, to be specified according to the doping profile. In spite of attaining negative values in certain regions in phase space (since Eq. (\ref{e3}) does not preserve the positive definiteness of $f$), the Wigner function can be used to evaluate ma\-cros\-co\-pic quantities like number, current and energy densities basically in the same manner as a classical probability distribution function. It is well-known \cite{Kluksdahl} that the Wigner-Poisson system is a convenient tool to describe the resonant tunneling diode (RTD). As the name says, the RTD relies on the quantum effect of tunneling, hence no classical model like the Vlasov-Poisson one could be appropriate for it. This is independent of the carrier concentration, although the Fermi-Dirac character needs to be taken into account in the drain region of $n^{+}nn^{+}$ diodes for instance. In applications, we have both FD and MB statistics when dealing with such nanoscopic systems \cite{Markowich}. 

Whether one call a charged particle system described by Eqs. (\ref{e3})-(\ref{e4}) a ``plasma" or ``an electron gas in a semiconductor" is just a matter of taste. At the end the basic equations are the same, perhaps with the difference that usually macroscopic plasmas 
consider an homogeneous ionic background. Actually the position-dependence of the doping $n_{0}(x)$ in Eq. (\ref{e4}) makes the analysis non\-linear {\it ab initio}, in contrast to the homogeneous case where one can start searching for linear waves and instabilities, just as in classical plasma theory. There is no profound reason to discard a complete analysis of the Wigner-Poisson and similar quantum kinetic models from the plasma physics point of view. Such a deleterious attitude would sound like an artificial censorship. Nevertheless, in certain si\-tua\-tions \cite{Asenjo} a clever order of magnitude analysis justify the neglect of the extra dispersion associated to the nonlocal term in Eq. (\ref{e3}), while keeping other quantum terms originating e.g. from the intrinsic spin dynamics. However, presently we are far from ha\-ving a complete picture of quantum plasma phenomena. Therefore a systematic neglect of quantum diffraction effects from the very beginning is not advisable. 

Equations (\ref{e3})-(\ref{e4}) constitute an integro-differential system, to be complemented with suitable initial and boundary conditions. Moreover, even if the Wigner-Poisson system is one of the simplest possible quantum kinetic models (non-relativistic,  electrostatic approximation, no spin effects etc.) by inspection of Eqs. (\ref{e3})-(\ref{e4}) it is clearly out of reach to access nonlinear phenomena in this framework, except perhaps from numerical simulation. This justifies the search for alternative models involving e.g. effective quantum potentials, time-dependent density functional theory or hydrodynamic equations. Each strategy has its merits and limitations. Focusing on hydrodynamic models, they can be deduced taking the moments of the quantum Vlasov equation (\ref{e3}), in the same way as the classical fluid equations can be derived from the moments of the Vlasov equation \cite{Nicholson}. The fact that the quantum Vlasov equation, as it stands, does not have collision terms is immaterial. The final result would be the same
after adding a suitable quantum Boltzmann-like collision integral in the right-hand side of Eq. (\ref{e3}), as long as this collision term is mass, momentum and energy preserving. In this respect notice that it is more direct to add phenomenolo\-gi\-cal dissipation terms in the fluid equations, if necessary. 

In the same way as in the transition from classical kinetic to classical hydrodynamic models, one faces a closure problem, since the evolution equation for the moment of order $N$ involves the moment of order $N+1$ of the Wigner function. One way to circumvent this difficulty is postulating an equation of state, compatible with the linear dispersion relation obtained from kinetic theory. However, it is evident that using fluid equations always imply a loss of information; the same holds in the classical scenario. Nevertheless, perhaps no one could seriously propose the understanding of the nonlinear structures in quantum plasmas based solely on numerical simulation of the kinetic equations. An interplay of kinetic theory and the analytic insights coming from the fluid approach is advisable. Note that quantum hydrodynamic models are ubiquitous e.g. in the analysis of ultra-small electronic devices and quantum chemistry \cite{Markowich}, seemingly always in the electrostatic approximation. 

To make these points more clear, we will explicitly write the quantum Vlasov equation for the Wigner function in the presence of electromagnetic fields assuming the Coulomb gauge, 
\begin{eqnarray}
&\strut& \frac{\partial f}{\partial t} + \frac{\bf p}{m}\cdot\nabla\,f + \frac{ie}{\hbar(2\pi\hbar)^3}\int\int d{\bf s}\,d{\bf p}'\,e^{\frac{i({\bf p}-{\bf p}')\cdot{\bf s}}{\hbar}}\times \nonumber  \\ &\times& \,[\phi({\bf r}+\frac{\bf s}{2},t)-\phi({\bf r}-\frac{\bf s}{2},t)]\,f({\bf r},{\bf p}',t)  \nonumber 
\\
&-& \frac{ie^2}{2\hbar m(2\pi\hbar)^3}\int\int d{\bf s}\,d{\bf p}'\,e^{\frac{i({\bf p}-{\bf p}')\cdot{\bf s}}{\hbar}} \times \nonumber \\ &\times& \,[A^{2}({\bf r}+\frac{\bf s}{2},t)-A^{2}({\bf r}-\frac{\bf s}{2},t)]\,f({\bf r},{\bf p}',t)  \nonumber \\
&+& \frac{e}{2m(2\pi\hbar)^3}\,\,\nabla\cdot\int\int d{\bf s}\,d{\bf p}'\,e^{\frac{i({\bf p}-{\bf p}')\cdot{\bf s}}{\hbar}} \times \nonumber \\ &\times& [{\bf A}({\bf r}+\frac{\bf s}{2},t)+{\bf A}
({\bf r}-\frac{\bf s}{2},t)]\,f({\bf r},{\bf p}',t) \nonumber \\
&-& \frac{ie}{\hbar m(2\pi\hbar)^3}\,\,{\bf p}\cdot\int\int d{\bf s}\,d{\bf p}'\,e^{\frac{i({\bf p}-{\bf p}')\cdot{\bf s}}{\hbar}} \times \nonumber  \\ \label{eqqq3} &\times& [{\bf A}({\bf r}+\frac{\bf s}{2},t)-{\bf A}({\bf r}-\frac{\bf s}{2},t)]\,f({\bf r},{\bf p}',t) = 0 \,, 
\end{eqnarray}
where $f = f({\bf r},{\bf p},t), \phi = \phi({\bf r},{\bf p},t)$ and introducing the vector potential ${\bf A} = {\bf A}({\bf r},t)$. Equation (\ref{eqqq3}) is coupled with Maxwell equations with self-consistent charge and current densities determined from the moments of the Wigner function. Since magnetic fields are an essential ingredient in plasmas, a quantum kinetic theory for plasmas should necessarily include Eq. (\ref{eqqq3}) or some 
variant of it. Obviously it is out of reach to progress from this pers\-pec\-ti\-ve, except for very simple cases (homogeneous 
magnetic fields, linear theory). Even the numerical simulation of the resulting quantum Vlasov-Maxwell system is clearly a 
challenge. To our knowledge there is no available 3D code for such a problem, which is still somehow limited (no relativistic nor spin effects, no exchange-correlation interactions taken into account). Facing this situation, one can: (i) postulate that magnetic fields does not exist in quantum plasmas and restrict forever to a peaceful electrostatic world; (ii) suggest clever approximations. 

Following the second alternative, one can for example proceed to quantum hydrodynamic models including magnetic fields. One may question the correct form or the applicability domain of such macroscopic formulations (as was detailed e.g. in \cite{Manfredi, Haas}); it is however absurd to question the need of simplified treatments, after open minded exa\-mi\-na\-tion of Eq. (\ref{eqqq3}) or any of its equivalents. In an optimistic view, one still could imagine that magnetic fields in phy\-si\-cal\-ly interesting situations could be treated in terms of purely classical models, letting quantum diffraction to the electric field alone. However, this would imply  significant error, as in the case of radio wave dispersion in a pulsar magnetosphere \cite{Melrose}, where we have the pa\-ra\-me\-ter $B/B_c$ of order unity, where $B$ is the magnetic field strength and $B_c = m^2 c^2/(e\hbar) \simeq 10^9 T$ is the critical magnetic field.

In close connection to magnetic field effects, one further quantum aspect deserving attention is the point (c) raised at the starting of this work: the dynamic of intrinsic magnetic moment of the charge carriers, or spin, see \cite{Brodin} for a review. The relevance of the spin dynamics should be evaluated in each particular instance. However, in this regard a representative dimensionless parameter is given by $\mu_B B/(\kappa_B T)$, which is the ratio between the Zeeman energy associated to the ambient magnetic field and the thermal energy, where $\mu_B = e\hbar/(2m)$ is the Bohr magneton. For a sufficiently high magnetic field (such as $B \simeq 10^{10} T$ in the vicinity of a magnetar), or for sufficiently low temperature, one can have a significant role of the spin dynamics for instance in the propagation of Alfv\'en waves \cite{Brodin}. 

To conclude, in this work we have shown that quantum plasma systems are ubiquitous in nature and briefly discussed their modeling. It is apparent that quantum plasma physics is an exciting, under construction area, in the need for creative approaches, both theoretical and experimental. We have not discussed the challenge of an efficient treatment of non-ideality (collisional) and re\-la\-ti\-vi\-ty terms. 

\vspace{.5cm}
\noindent
{\bf Acknowledgments}\\
\noindent
This work was supported by CNPq (Conselho Nacional de Desenvolvimento Cient\'{\i}fico e Tecnol\'ogico). We also acknowledge 
Prof. Padma Kant Shukla for his continuous support.

\end{document}